\documentstyle[aps,prl,psfig]{revtex}
\begin{document}
\topmargin=0.0cm

\twocolumn

\title {Influence of a knot on the strength of a polymer strand}

\author{A. Marco Saitta$^*$, Paul D. Soper$^{\dag}$, 
E. Wasserman$^{\dag}$, and Michael L. Klein$^*$}

\address{$^*$Center for Molecular Modeling, Dept. of Chemistry,
University of Pennsylvania, Philadelphia, PA 19104-6202, USA\\
$^{\dag}$DuPont Central Research and Development, Expt. Station,
Wilmington, DE 19880-0328, USA}

\date{\today}
\maketitle

{\bf

Many experiments have
been done to determine the relative strength of different knots,
and these show that the break in a
knotted rope almost invariably occurs at a point just outside the `entrance'
to the knot$^1$.
The influence of knots on the properties of polymers has become of great
interest, in part because of their effect on mechanical properties$^2$.
Knot theory$^{3,4}$ applied to the topology of macromolecules$^{5-8}$
indicates that the simple
trefoil or `overhand' knot is likely to be present with high probability in
any long polymer strand$^{9-12}$.
Fragments of DNA have been observed to contain such
knots in experiments$^{13,14}$ and computer simulations$^{15}$.
Here we use {\it ab initio} computational methods$^{16}$ to 
investigate the effect of a trefoil knot on the breaking strength
of a polymer strand. We find that the knot weakens the strand
significantly, and that, like a knotted rope, it breaks under
tension at the entrance to the knot.
}

Little is known about the structure and properties of knots at the atomic
level. For example, the minimum number of carbon atoms that can be sustained
as a trefoil in a polyethylene strand is not well established$^{5,19}$.
Polyethylene is the simplest polymer and therefore an excellent generic
system to study the fundamental properties of a knotted chain.
(Rigorously, a knot is a closed loop; but here, as in ref. 1,
we apply the term to a strand with unlinked ends.
As a representative polyethylene-like system we chose the linear molecule
$n$-decane (${\rm C_{10}H_{22}}$).
Empirical models that utilize force constants 
for bond stretch, bend and torsion$^{18-20}$,
although useful for studying bulk properties of chain molecules, are
not applicable to the case of chain rupture.
First-principles calculations$^{16}$, on the other hand,
have been
shown to yield satisfactory results for structural and mechanical properties
of hydrocarbon-based polymeric systems$^{21,22}$.
The {\it ab initio} equilibrium structural parameters, after
optimization, are in excellent agreement with experiment, with errors in
bond lengths and angles smaller than 1\% (ref. 22). 
We studied the response of the $n$-decane
molecule to uniaxial strain by systematically increasing the
separation $L$ between the terminal carbon atoms. For a fixed value of this
separation the system was heated to room temperature and then cooled via a
simulated annealing technique. The decane molecule remained intact for an
elongation up to 18.5\% beyond its equilibrium length. Deviations of the
$\rm C_1$-$\rm C_2$ ad $\rm C_9$-$\rm C_{10}$ bond lengths
from the mean value increase significant with the applied strain. 
Under maximum loading the terminal bonds become stretched 
to $\sim$ 1.8$\AA$ compared with $\sim$ 1.7$\AA$ for the others 
and the terminal $\rm C$-$\rm C$-$\rm C$ bond angles increase to $\sim$ 135 deg.

\begin{figure}
\centerline{\psfig{figure=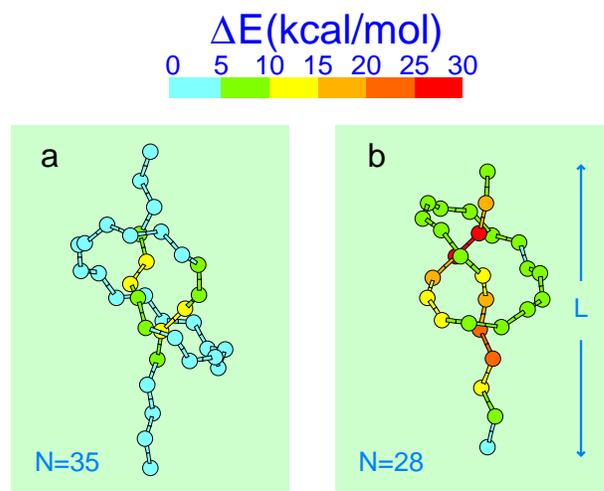,width=8.0truecm}}
\label{fig1}
\caption{Strain energy distribution in a knotted polymer strand.
Shown are distributions in chains of
35 ({\bf a}) and 28 ({\bf b}) carbon atoms
taken from constrained classical MD simulations.
When the knot
is sufficiently tightened, the strain energy localizes
mostly on the bonds immediately outside its entrance points.}
\end{figure}

For larger elongation the molecule dissociates into two radicals with
$\rm C_{10}H_{22} \rightarrow \rm C_9H_{19}{\bf\cdot} + CH_3{\bf\cdot}$.
The bond that dissociates involves
(randomly) one of the two atoms where the tension is applied.
We obtain 83 kcal mol$^{-1}$ for the dissociation energy for the methyl group,
which compares favourably with the experimental dissociation
enthalpy of $\sim$ 87 kcal mol$^{-1}$. A similar calculation performed for $n$-undecane,
${\rm C_{11}H_{24}}$, but with the tension still applied 
to the $\rm C_1$ and $\rm C_{10}$ atoms, shows
that the $\rm C_9$-$\rm C_{10}$ bond breaks to yield an ethyl radical. The
calculated dissociation energy in this case is 81 kcal mol$^{-1}$ compared with
the experimental enthalpy of $\sim$ 82 kcal mol$^{-1}$. As in the previous
case, the
hybridization of the two key carbon atoms changes from the initial
tetrahedral $sp^3$ to planar $sp^2$ on formation of the product radicals.

To generate a starting configuration for the {\it ab initio} study 
we first performed
classical molecular dynamics (MD) calculations on a 
loosely knotted polyethylene-like alkane chain, with $N$=144 carbon atoms,
in which a trefoil was introduced by adding an appropriate set of {\it gauche}
defects$^{17}$.
The potential model employed was based on the united atom scheme$^{18,19}$ but
with {\it ab initio} intra-chain stretch and bend force constants 
taken from the present study.
MD calculations were carried out on the
knotted chain for a sequence of increasing (constrained) end-to-end
distances, $L$, which gradually resulted in a tighter knot. For any given
$L$, 200 ps of room temperature dynamics,
followed by 50 ps of slow cooling, was utilized to
relax the system in the trefoil configuration$^{23}$.
(This run length 
seemed appropriate given that even the slowest
longitudinal phonons can propagate along the entire length of the molecule
about 1,000 times during this trajectory.)

\begin{figure}
\centerline{\psfig{figure=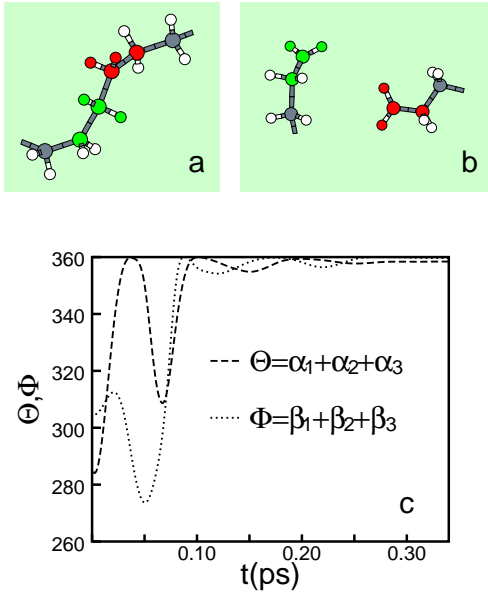,width=7.0truecm}}
\label{fig2}
\caption{
Analysis of chain rupture.
The breakpoint in a $N$=28 trefoil before ({\bf a}) and after ({\bf b})
chain dissociation, with $\rm C_{22}$ and $\rm C_{23}$ and their
neighbouring atoms highlighted in green and red respectively.
Panel {\bf c} shows the time evolution of the sum of the
respective bond angles, which reveals an asymptotic $sp^2$ 
hybridization of the carbons involved in the break.
The calculations are based on density functional theory
with Becke and Lee-Yang-Parr (BLYP)$^{24,25}$ corrections to the exchange and
correlation functionals, respectively. Electronic spin polarization
has been included within the local spin-density approximation scheme.
A valence electron pseudopotential
scheme is employed with a plane-wave kinetic-energy cutoff at 60 Ry and a
$\Gamma$-point sampling of the Brillouin zone. The simulation supercell is large
enough to avoid interactions with periodic images.}
\end{figure}

The global (either constrained or unconstrained) minimum corresponds to an
unknotted polymer, but the migration barrier for the knot to propagate along
the chain, and thus the disentanglement time, increases dramatically as the
trefoil tightens. Thus even at room temperature the system
is `trapped' in the knotted configuration for the timescale of the
calculations. Under these circumstances, the portions of the chain
far from the trefoil, although extremely important in disentanglement processes,
are likely to have little or no effect on the behavior of the knot under stress.
Accordingly, as the knot was tightened, atoms far from it
were removed from the chain, in order to eventually make the system
tractable with first-principles methods.

For each classical MD calculation, we monitored the strain distribution
along the chain.
Even for $N=50$, the chain with a trefoil stores a rather small amount of
strain energy. However, when the chain was shortened to $N$=35 a
pronounced inhomogeneity appeared in the strain energy distribution (Fig. 1a). 
The shape of the strain energy distribution is essentially
the same, at a given $L$, independent of the details of
the simulation, such as the initial configuration and temperature or the
length of the MD simulation or a reasonable change in the interaction
potentials.
(The potentials that we have employed in the MD simulations 
were all fitted to the properties of alkane molecules close to their
equilibrium structures. Thus, in the present case where we are dealing with
highly distorted systems, the model's force constants are likely to yield only
semi-quantitative results.)
Further shortening of the chain to $N$=28 produced a trefoil
(Fig. 1b) with distortions in bond lengths and angles still below the critical
values that yield bond breaking in the linear alkanes.

The constrained equilibrium positions obtained from the 250 ps of 
classical MD evolution of the ${\rm C_{28}}$ $n$-alkane 
provided the carbon skeleton of the starting configuration 
for the first-principles study of the ${\rm C_{28}H_{58}}$ molecule. 
A series of Car-Parrinello MD (CPMD) simulations$^{16,23-25}$ were carried out
with the separation between $\rm C_1$ and $\rm C_{28}$ fixed at distances $L$
between 11.50 and $14.00 \AA$.
In each case, the system was allowed to evolve dynamically at room
temperature, after which it was cooled.
Typically, during the dynamical evolution individual
bonds undergo large amplitude thermal fluctuations, which decrease on
cooling.

During the room temperature $L =13.50 \AA$ CPMD run for $N$=28, dissociation
occurred at a bond
location, just outside the entrance to the knot, well separated from the
terminal atom where the tension was applied. Figure 2 shows the region of
the knot where the chain actually ruptured, both before (Fig. 2a) and after
(Fig. 2b) the break.
The change in hybridization of the $\rm C_{22}$ and $\rm C_{23}$ atoms
from $sp^3$ to $sp^2$ is clearly evident also in Fig. 2c, where
we display the time evolution of the system immediately before
and after the break. Figure 3 shows a
snapshot of the electronic charge density after the break occurred.
The gap in the charge density between the $\rm C_{22}$ and $\rm C_{23}$ atoms
confirms the bond breaking at this position.
Other simulations carried out with $L \geq 13.50 \AA$ yield
breakpoints at one of the two bonds coming out from the knot$^{25}$.
(Density-functional energy barriers are not quantitatively very accurate,
but the general qualitative description of phenomena, as well as the
dissociation energies involved in such processes, are known to be
quite reliable.)

\begin{figure}
\centerline{\psfig{figure=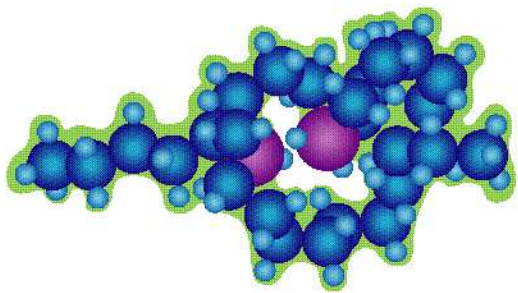,width=8.0truecm}}
\label{fig3}
\caption{
Schematic electron charge density immediately after the break.
Carbon and hydrogen atoms are displayed with spheres corresponding
to their respective covalent radii.
The green area is a contour plot of the region containing 87\%
of the total electronic charge.
A gap in the charge density, and thus bond-breaking,
is observable between the two highlighted $\rm C$ atoms.}
\end{figure}

The first-principles CPMD calculations provide the total strain
energy but not its profile along the chain. 
The latter can be obtained by making use of an
{\it ab initio} force constants model, which is able to reproduce
the total strain energy to $\sim$ 20\%. Figure 4 shows
the evolution of the strain energy distribution during structural
relaxation of a $N$=30 trefoil just before breaking occurs.
The strain energy is mainly localized in the two
symmetric bonds that are outside the entrance to the knot. Our calculations
suggest that 23 carbon atoms form the tightest knot that can be sustained in
a polyethylene strand without it breaking. 

The strain energy stored in the knot at breaking point is 12.7 kcal
mol$^{-1}$ per
C-C bond, which is considerably smaller than the value 16.2 kcal
mol$^{-1}$ for the
linear unknotted case. Thus, we find that the presence of the knot has
significantly weakened the strand in which it is tied.

While the dynamical evolution of the present constrained {\it ab initio} MD
simulations was sufficient to observe bond breaking, no attempt was made to
allow for recombination of the resulting radicals or reactions of the
radicals with other parts of the chain. The study of these effects as well
as the role of chain branching and the influence of neighbouring chains is
left for future research. These factors will likely provide a deeper
understanding of how the interplay between inter- and intra-molecular
effects contributes to the mechanical properties 
of real polymer samples$^2$.

{\it Note added in proof}: Arai {\it et al.}$^{26}$ have recently
reported the knotting of actin filaments and DNA molecules using
optical tweezers. They find that the breaking stress for actin is
significantly lower than that of the unknotted filaments, and that
the breakage point is at the entrance to the knot, as our calculations
predict.

\begin{figure}
\centerline{\psfig{figure=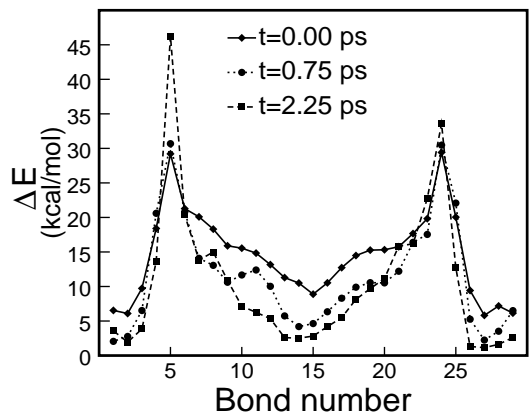,width=8.0truecm}}
\label{fig4}
\caption{
Relaxation of the strain energy distribution.
Shown is the distribution along a $\rm C_{30}H_{62}$ chain containing
a trefoil during CPMD energy minimization at $T=0 K$.
The $t=0$ configuration is an average of the equilibrium positions
obtained along classical MD simulations performed with different
initial states and temperatures.
The segments outside the knot and its central portion
relax, while stress tends to concentrate mostly on the two 
entrance bonds (see Fig.1).}
\end{figure}

\noindent
\rule{8.65cm}{0.1mm}

\scriptsize

\noindent
1. Ashley, C.W. {\rm The Ashley Book of Knots} (Doubleday, New York, 1993).

\noindent
2. Bayer, R.K. Structure transfer from a polymeric melt to the solid state Part III: Influence of knots on structure and mechanical properties of semicrystalline polymers. {\it Colloid Polym. Sci.} {\bf 272}, 910-932 (1994).

\noindent
3. Atiyha, M.F. {\rm The geometry and physics of knots} (Cambridge University Press, 1990).

\noindent
4. Katritch., V. {\it et al.} Geometry and physics of knots.
{\it Nature} {\bf 384}, 142-145 (1996); Katritch., V. {\it et al.}
Properties of ideal composite knots. {\it Nature} {\bf 388}, 148-151
(1997).

\noindent
5. Frisch, H.L. \& Wasserman, E. Chemical topology. {\it J. Am. Chem. Soc.} {\bf 83}, 3789-3795 (1961).

\noindent
6. Mislow, K. {\rm Introduction to stereochemistry} (W.A. Benjamin, New York, 1965).

\noindent
7. Schill, G. {\rm Catenanes, Rotaxanes, and Knots} (Academic, New York, 1971).

\noindent
8. Walba, D.M. Topological stereochemistry. {\it Tetrahedron} {\bf 41}, 3161-3212 (1985).

\noindent
9. Frank-Kamenetskii, M.D., Lukashin, A.V. \& Vologodskii, A.V. Statistical mechanics and topology of polymer chains. {\it Nature} {\bf 258}, 398-402 (1975).

\noindent
10. Frisch, H.L. Macromolecular topology - Metastable isomers from pseudo interpenetrating polymer network. {\it New J. Chem.} {\bf 17}, 697-701 (1993).

\noindent
11. Mansfield, M.L. Knots in hamiltonian cycles. {\it Macromolecules} {\bf 27}, 5924-5926 (1994).

\noindent
12. van Rensburg, E.J., Sumners, D.A.W., Wasserman, E. \& Whittington, S.G. Entanglement complexity of self-avoiding walks. {\it J. Phys. A.} {\bf 25}, 6557-6566 (1992).

\noindent
13. Wasserman, S.A. \& Cozzarelli, N.R. Biochemical topology: applications to DNA recombination and replication. {\it Science} {\bf 232}, 951-960 (1986).

\noindent
14. Shaw, S.Y. \& Wang, J.C. Knotting of a DNA chain during ring closure. {\it
Science} {\bf 260}, 533-536 (1993).

\noindent
15. Schlick, T. \& Olson, W.K. Trefoil knotting revealed by molecular
dynamics simulations of supercoiled DNA. {\it Science} {\bf 257},
1110-1115 (1992).

\noindent
16. Car, R. \& Parrinello, M. Unified approach for molecular dynamics and density-functional theory. {\it Phys. Rev. Lett.} {\bf 55}, 2471-2474 (1985).

\noindent
17. de Gennes, P.-G. Tight knots. {\it Macromolecules} {\bf 17}, 703-704 (1984).

\noindent
18. Siepmann, J.I., Karaborni, S. \& Smit, B. Simulating the critical-behavior of complex fluids. {\it Nature} {\bf 365}, 330-332 (1993).

\noindent
19. Mundy, C.J., Balasubramanian, S., Bagchi, K., Siepmann, J.I. \& Klein, M.L. Equilibrium and non-equilibrium simulation studies of fluid alkanes in bulk and at interfaces. {\it Faraday Discuss.} {\bf 104}, 17-36 (1996).

\noindent
20. Karasawa, N., Dasgupta, S. \& Goddard, W.A. III Mechanical properties and force-field parameters for polyethylene crystal. {\it J. Phys. Chem. US} {\bf 95}, 2260-2272 (1991).

\noindent
21. Ancilotto, F., Chiarotti, G.L., Scandolo S. \& Tosatti E. Dissociation of methane into hydrocarbons at extreme (planetary) pressure and temperature. {\it Science} {\bf 275}, 1288-1290 (1997).

\noindent
22. Montanari, B. \& Jones, R.O. Density functional study of crystalline polyethylene. {\it Chem. Phys. Lett.} {\bf 272}, 347-352 (1997), and references therein.

%
%
%

\noindent
23. Martyna, G.J. {it et al.} PINY Code. {\it Comp. Phys. Comm.} (in the
press).

\noindent
24. Becke, A.D. Density-functional exchange-energy approximation with correct asymptotic behavior. {\it Phys. Rev. A} {\bf 38}, 3098-3100 (1988)

\noindent
25. Lee, C., Yang, W. \& Parr, R.G. Development of the Colle-Salvetti correlation-energy formula into a functional of the electron density. {\it Phys. Rev. B} {\bf 37}, 785-789 (1988).

\noindent
26. Arai, T. {\it et al.} Tying a molecular knot with optical tweezers.
{\it Nature} (in the press).

\centerline{~}
\tiny
\noindent
Acknowledgements. The work was supported in part by the
National Science Foundation.
\centerline{~}

\noindent
Correspondence~and~request~for~material~should be addressed to M.L.K.
(e-mail:~klein@lrsm.upenn.edu).\\
\rule{8.65cm}{0.1mm}

\footnotesize

\end{document}